\documentclass[10pt,aps,prl,twocolumn,preprint,nofootinbib]{revtex4}

\usepackage{amsmath}    
\usepackage{graphicx}   
\usepackage{color}
\usepackage{verbatim}

\usepackage{cancel}

\begin{document}

\title{Magnetic Field Expulsion from an Infinite Cylindrical Superconductor}
\author{Miguel C. N. Fiolhais}
\email{miguel.fiolhais@cern.ch}   
\affiliation{LIP, Department of Physics, University of Coimbra, 3004-516 Coimbra, Portugal}
\author{Hanno Ess\'en}
\email{hanno@mech.kth.se}   
\affiliation{Department of Mechanics, KTH, 10044 Stockholm, Sweden}

\date{\today}

\begin{abstract}

The solutions of the London equations for the magnetic field expulsion from superconductors are presented in this paper for the cylindrical symmetry.
The result is analyzed in detail and represented numerically for the case of a uniform external magnetic field in the transverse plane. In particular, several contour 
plots of the magnetic energy density are depicted for the regions inside and around the superconducting area for a wide range of penetration lengths, showing how the expulsion and penetration of the magnetic field evolve with the ratio between the penetration length and the cylinder radius.
\\
\\
The following paper is published in Physica C: http://www.sciencedirect.com/science/article/pii/S0921453413004541

\end{abstract}

\maketitle

\section{Introduction}
\label{intro}

The first phenomenological description of the magnetic field expulsion from superconductors appeared soon after the discovery of the Meissner-Ochsenfeld effect~\cite{meissner}, by the London brothers in 1935~\cite{london}. Ever since, several discussions have been presented along the years on the nature of the magnetic field expulsion configuration in the scientific literature~\cite{cullwick, degennes,pfleiderer,karlsson,badiamajos,kudinov,zhilichev,matute,prozorov,ejrnaes,fiolhais,essen,essen2,fiolhais2,fiolhais3}. These studies comprise both static and time-dependent analyses, several different geometries, and even the case of non-homogenous superconductors. However, while the spherical case is studied in~\cite{matute,reitz,batygin}, and the cylinder case in a parallel magnetic field is presented in~\cite{batygin}, to the best of our knowledge the detailed solution of a superconducting cylinder in a transversal magnetic field cannot be easily found in the academic literature.
This very same case has been previously addressed by Zhilichev \cite{zhilichev} by proposing a macroscopic shell model, but without providing an explicit solution for the London condition. To fill this gap, we derive the solution of the London equations in this paper, for the magnetic field expulsion from an infinite homogeneous superconducting cylinder in a constant external magnetic field in the transverse plane. The dependence of the strength and configuration of the expelled magnetic field with the London penetration depth is examined in detail.

The London equations for the magnetic field flux expulsion out of a superconductor can be expressed, in terms of the magnetic vector potential $\mathbf{A}$, in a single equation:
\begin{equation}
\mathbf{j} = - \frac{1}{\lambda^2} \mathbf{A} \, ,
\end{equation}
where $\mathbf{j}$ represents the electric current and $\lambda$ the London penetration depth. After applying the Maxwell's equation for the static case, and by taking the Coulomb gauge ($\nabla \cdot \mathbf{A} = 0$), one obtains:
\begin{equation}
\nabla^2 \mathbf{A}  - \frac{1}{\lambda^2} \mathbf{A} = 0 \, ,
\label{eq:londonA}
\end{equation}
or, in terms of the magnetic field,
\begin{equation}
\nabla^2 \mathbf{B}  - \frac{1}{\lambda^2} \mathbf{B} = 0 \, .
\end{equation}
This equation is perhaps the most common form of the London equation for the magnetic field expulsion inside the superconducting region, 
and, in the cartesian coordinate system, corresponds to the well-known Helmholtz's equation for a complex wavenumber, for each of the vector components. The magnetic field is, therefore, suppressed in the interior region through negative exponential dependencies, or, in the cylindrical and spherical symmetries, through modified Bessel functions of the first kind~\cite{abramowitz}.

\section{Cylindrical Symmetry}

Consider a cylindrical superconductor with radius $R$ in an external constant perpendicular magnetic field, in such a way the external magnetic field points in the $y$-direction 
and the cylinder axis coincides with the $z$-axis. In order to obtain the final static vector potential configuration resulting from the magnetic flux expulsion, and assuming the Coulomb gauge, one must solve the Maxwell equation for the outer region:
\begin{equation}
\nabla \times \mathbf{B} = \nabla \times \left ( \nabla \times \mathbf{A} \right ) = \nabla^2 \mathbf{A} = 0 \, ,
\label{eq:outer}
\end{equation}
and the London equation (\ref{eq:londonA}) for the inner region.
Due to the symmetry of the system, the magnetic field has no component in the $z$-axis direction,
\begin{equation}
\mathbf{B} = \frac{1}{\rho}\frac{\partial A_z}{\partial \phi} \hat{\rho} - \frac{\partial A_z}{\partial \rho} \hat{\phi} \, ,
\end{equation}
and, therefore, the axial component of the vector potential $A_z$ is the only relevant component needed to compute the magnetic field.
For the outer region, this component can be determined from equation (\ref{eq:outer}):
\begin{equation}
\left ( \frac{1}{\rho^2}\frac{\partial^2 A_z}{\partial \phi^2} + \frac{\partial^2 A_z}{\partial \rho^2} + \frac{1}{\rho}\frac{\partial A_z}{\partial \rho} \right ) \hat{z}  = 0 \, ,
\end{equation}
which is simply the Laplace equation for $A_z$. The most general solution in the outer region is, therefore,
\begin{eqnarray}
A_z  =  & \sum_{n=0}^{+\infty} &  \left [ A_n \cos \left ( n\phi \right ) + B_n \sin \left ( n\phi \right ) \right ] \nonumber \\ 
&\times & \left [ C_n \rho^n + D_n \rho^{-n} \right ] \, .
\end{eqnarray}
Assuming the magnetic field takes the form of the external field in the limit $\rho \rightarrow + \infty$,
\begin{equation}
\mathbf{B}_0 = B_0 \mathbf{\hat{y}} =  B_0 \left ( \mathbf{\hat{\phi}} \cos \phi +  \mathbf{\hat{\rho}} \sin \phi \right ) \, , 
\end{equation}
and in order to respect this boundary condition, the vector potential becomes,
\begin{equation}
A_z = \textrm{const.} + A_1 \left ( C_1 \rho + D_1 \rho^{-1} \right ) \cos \phi \, ,
\end{equation}
where $A_1 C_1 = - B_0$. The remaining parameters must be determined by the boundary conditions at the surface of the superconductor.

The general solution of the London equation in the interior region, for the axial component of the vector potential, is somewhat more complicated:
\begin{eqnarray}
A_z = & \sum_{n=0}^{+\infty} & \left [ E_n \cos \left ( n\phi \right ) + F_n \sin \left ( n\phi \right ) \right ] \nonumber \\
& \times & \left [ G_n I_{n} (\rho/\lambda) + H_n K_{n} (\rho/\lambda) \right ] \, ,
\end{eqnarray}
where $I_{n} (\rho/\lambda)$ and $K_{n} (\rho/\lambda)$ are the cylindrical modified Bessel functions of the first and second kind, respectively. As the modified Bessel functions of the second kind diverge at $\rho=0$, we can exclude them \emph{a priori}.
 
To ensure continuity at the superconducting surface, the $A_z$ function must take the following form on the inner region,
\begin{equation}
A_z = E_0 G_0 I_{0} (\rho/\lambda) + E_1 G_1 I_{1} (\rho/\lambda) \cos \phi  \, ,
\end{equation}
where,
\begin{equation}
 E_1 G_1 I_{1} (R/\lambda) = - B_0 R + A_1 D_1 / R \, .
\end{equation}
Since the electric current is distributed in volume, \mbox{$\mathbf{j} = - \frac{1}{\lambda^2} \mathbf{A}$}, there are no pure surface currents at the cylindrical surface, 
allowing some penetration of the magnetic field. Therefore, the following relation can be used as a boundary condition:
\begin{equation}
\mathbf{k} = \mathbf{\hat{n}} \times \left ( \mathbf{B}^+ - \mathbf{B}^- \right )  = 0 \, ,
\end{equation}
where $\mathbf{k}$ is the surface current, $\mathbf{\hat{n}}$ is the unitary vector normal to the surface and, $\mathbf{B}^+$ and $\mathbf{B}^-$ are the magnetic field at the cylinder surface in the outer and inner limits, respectively. As a result, one can establish the following condition:
\begin{equation}
\frac{\partial A_z}{\partial \rho^+} = \frac{\partial A_z}{\partial \rho^-} \, ,
\end{equation}
which, with the use of the relation,
\begin{equation}
\frac{\partial I_{\nu}(x)}{\partial x} = I_{\nu-1} (x) - \frac{\nu}{x} I_{\nu} (x)
\end{equation}
in the differentiation process, leads to,
\begin{eqnarray}
\mathbf{B}_{\rho > R} & = & B_0 \left ( 1 -  \frac{R^2}{\rho^2} + 2 \lambda \frac{R}{\rho^2} \frac{I_{1} (R/\lambda)}{I_{0} (R/\lambda)} \right ) \sin \phi \, \hat{\rho} \nonumber \\
& + &  B_0 \left ( 1 +  \frac{R^2}{\rho^2} - 2 \lambda \frac{R}{\rho^2} \frac{I_{1} (R/\lambda)}{I_{0} (R/\lambda)} \right ) \cos \phi \, \hat{\phi} \, , \nonumber \\
\mathbf{B}_{\rho < R} & = & 2 B_0 \frac{\lambda}{\rho} \frac{I_{1} (\rho/\lambda)}{I_{0} (R/\lambda)} \sin \phi \, \hat{\rho} \nonumber \\
& + & 2 B_0  \left ( \frac{I_{0} (\rho/\lambda)}{I_{0} (R/\lambda)} - \frac{\lambda}{\rho}  \frac{I_{1} (\rho/\lambda)}{I_{0} (R/\lambda)}  \right ) \cos \phi \, \hat{\phi}  \, .
\label{magneticfield}
\end{eqnarray}
In summary, the magnetic field vector was easily determined by imposing that the axial component of the vector potential must a differentiable function in any point of space, in particular, at the superconductor surface. A detailed analysis of these solutions is presented in the next section.

\section{Discussion of Results}

\begin{figure}[!ht]
\begin{center}
\includegraphics[height=3.8cm]{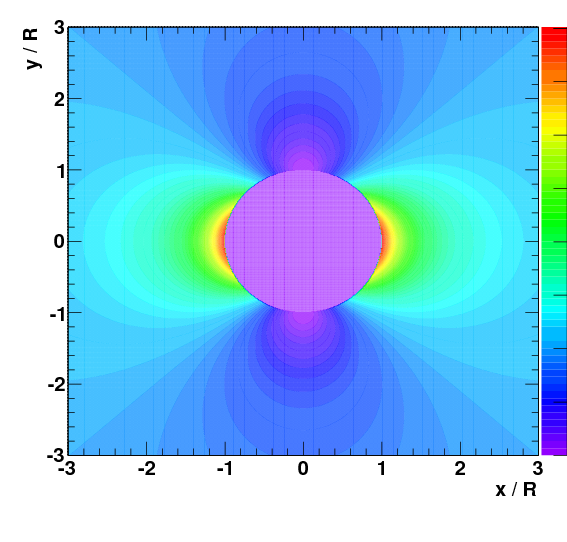} \includegraphics[height=3.8cm]{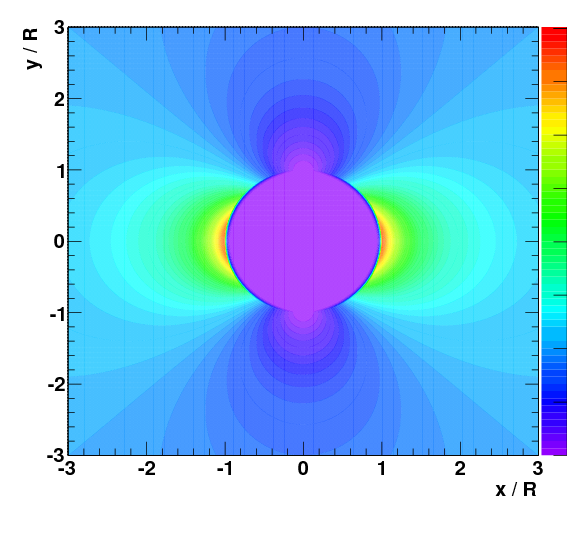} \\ \includegraphics[height=3.8cm]{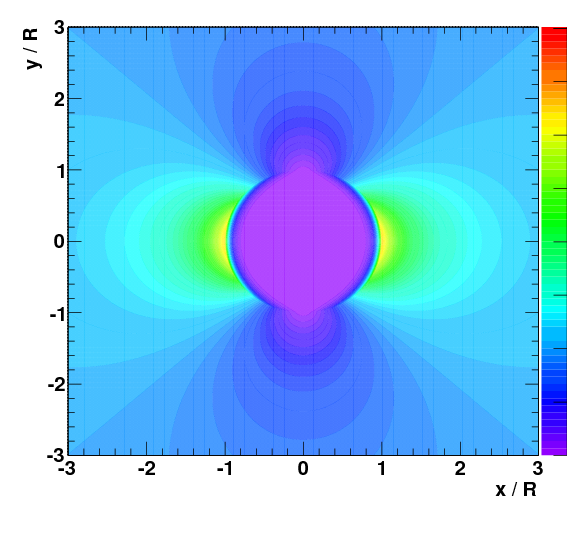} 
\includegraphics[height=3.8cm]{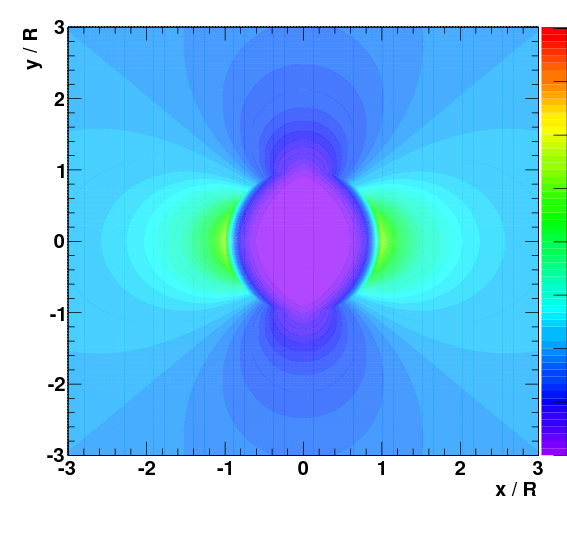} \\ \includegraphics[height=3.8cm]{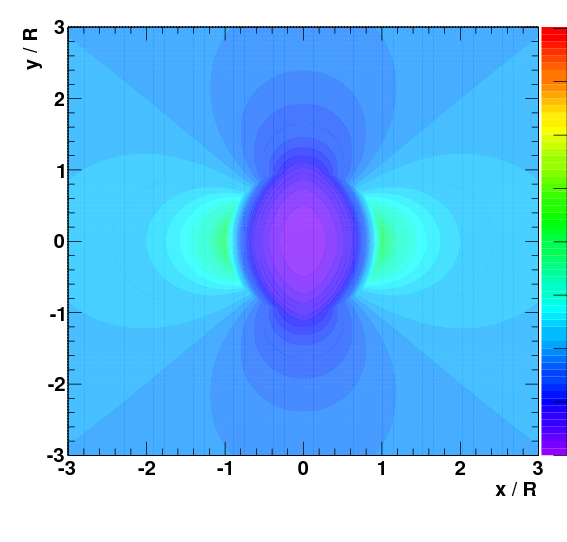} \includegraphics[height=3.8cm]{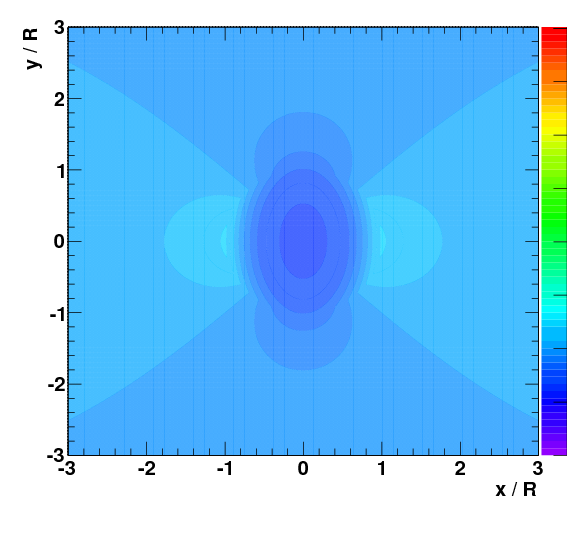}
\caption{Contour plots of the magnetic energy density for $\lambda = 0$ (top left), $\lambda = 0.03~R$ (top right), $\lambda = 0.1~R$ (center left), $\lambda = 0.2~R$ (center right), $\lambda = 0.4~R$ (bottom left), and $\lambda = R$ (bottom right).}
\label{contour}
\end{center}
\end{figure}

Several plots of magnetic field energy density are shown in Figure~\ref{contour}, for different values of the London penetration depth. 
Each of the pictures represents a cross section of the cylinder in the x-y plane ranging from minus 3 to 3 in steps of 0.01 radius units for both x and y directions. The magnetic energy density was computed from equations (\ref{magneticfield}) in units of $B_0^2$, \emph{i.e.} the magnetic energy density at an infinite distance. It is clearly visible that the purple circle, 
corresponding to field expulsion, shrinks to an ellipsoid as the London depth increases and the magnetic field penetrates more and more.
The magnetic field energy density reaches its maximum value at the cylinder's border in the horizontal axis due to the concentration of field lines in this region, 
and the minimal region is located at the top and bottom of the cylinder. As the penetration depth increases, the maximal regions tend to fade while entering the cylinder 
to form the ellipsoidal shaped region. 

\begin{figure}[!t]
\begin{center}
\includegraphics[height=5.5cm]{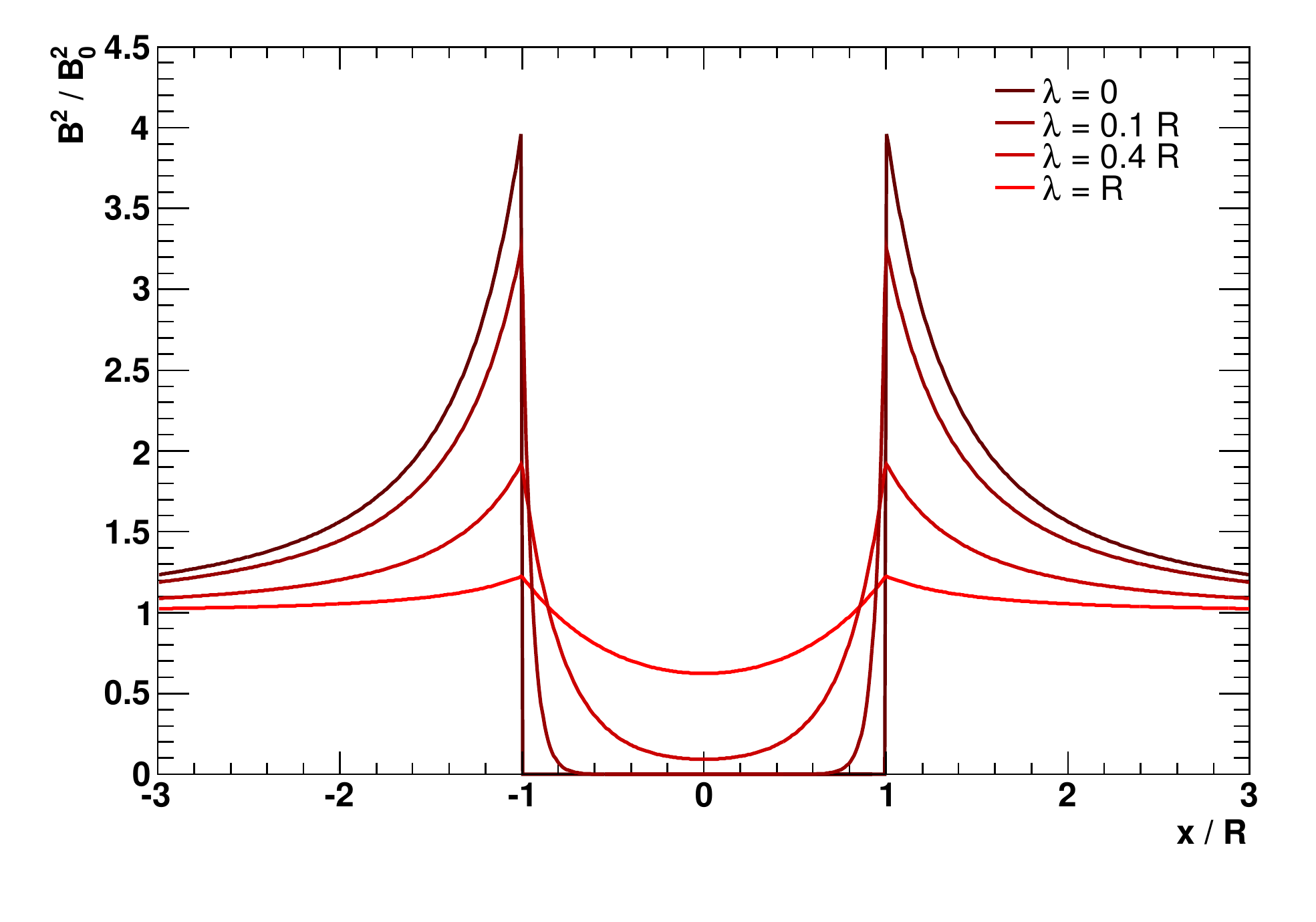}
\caption{Dependence of the magnetic energy density with the $x$-coordinate at $y=0$ for different penetration lengths.}
\label{onedim}
\end{center}
\end{figure}

Furthermore, one can see that no dramatic changes are perceptible until the penetration depth becomes one order of magnitude lower than the radius of the cylinder, when the ellipsoid starts to form. It is also worth noting that when the London depth equals the radius of the cylinder the magnetic field expulsion is almost imperceptible, or in other words, the superconductor becomes transparent to the magnetic field. This is also visible in Figure~\ref{onedim}, where the dependence of the magnetic energy density along the $x$-axis is shown at $y = 0$ for different penetration depths. In particular, a single jump on the order of magnitude, from $\lambda = 0.1~R$ to $\lambda = R$, is enough to have a significant phenomenological impact, \emph{i.e.} the difference between an almost fully expelled field and a considerable field penetration.

Finally, it is also worth discussing the range of applicability of these solutions in both type-I and type-II superconductors. In type-I superconductors, the London penetration length is of the same order of magnitude or smaller than the coherence length~\cite{ginzburg,fiolhais4}. Therefore, the most accurate description is not given by the London equations alone, but by the Pippard's model instead~\cite{pippard}. However, in type-II superconductors, the London condition indeed provides a very good description of the electromagnetic field inside the superconductor in the Meissner state, \emph{i.e.} below the lower critical field. Above the lower critical field, the superconductor enters a mixed state, where the magnetic field is allowed to penetrate the superconducting region through quantized vortices~\cite{abrikosov}, and the London solutions are no longer an accurate description.

\section{Conclusions}

We presented a detailed solution of the London equation for the magnetic field expulsion from an infinite cylindrical superconductor, which can 
be easily obtained by using physical and differentiability arguments at the boundary conditions. The field expulsion is analyzed  
with the use of contour plots of the magnetic energy density to show how it evolves with the value of the penetration length. 
In this case, the magnetic field penetrates through the formation of an elliptic region which becomes more pronounced as the penetration depth increases.


\end{document}